\documentclass[12pt]{article}

\usepackage{graphicx}
\usepackage{hyperref}
\usepackage{cite}
\usepackage{fixltx2e}

\newcommand\pubnumber{}
\newcommand\pubdate{\today}

\newcommand{\ordEW}[1]{$\mathcal{O}(\alpha^{#1})$}
\newcommand{\ordQCDEW}[2]{$\mathcal{O}(\alpha_s^{#1} \alpha^{#2})$}
\newcommand{\NLO}[1]{\textsc{NLO}\textsubscript{#1}}

\def\institute{MTA-DE Particle Physics Research Group \\ H-4002 Debrecen, PO Box 400, Hungary }
\def\support{\footnote{Work supported by grant K 125105 of the National Research, Development and Innovation Office in Hungary.}}

\def\Title#1{\begin{center} {\Large #1 } \end{center}}
\def\Author#1{\begin{center}{ \sc #1} \end{center}}
\def\Address#1{\begin{center}{ \it #1} \end{center}}

\newcommand\pubblock{\rightline{\begin{tabular}{l} \pubnumber\\
         \pubdate  \end{tabular}}}
\newenvironment{Abstract}{\begin{quotation}  }{\end{quotation}}
\newenvironment{Presented}{\begin{quotation} \begin{center} 
             PRESENTED AT\end{center}\bigskip 
      \begin{center}\begin{large}}{\end{large}\end{center} \end{quotation}}

\def\beq{\begin{equation}}
\def\eeq#1{\label{#1}\end{equation}}
\def\eeqn{\end{equation}}

\def\beqa{\begin{eqnarray}}
\def\eeqa#1{\label{#1}\end{eqnarray}}
\def\eeqan{\end{eqnarray}}

\let\bar=\overbar

\def\Dslash{\not{\hbox{\kern-4pt $D$}}}
\def\dslash{\not{\hbox{\kern-2pt $\del$}}}

\def\msb{{\bar{\ssstyle M \kern -1pt S}}}

\begin{document}

\begin{titlepage}

\pubblock

\vfill
\Title{$t\bar{t}W^\pm$ at NLO accuracy with realistic final states}
\vfill
\Author{ Giuseppe Bevilacqua\support}
\Address{\institute}
\vfill
\begin{Abstract}
In this proceedings we summarize state-of-the-art theoretical predictions for the process of $t\bar{t}W^\pm$ production at the LHC. After a review of the status of inclusive calculations, we discuss recent advancements in exclusive predictions.
\end{Abstract}
\vfill
\begin{Presented}
$13^\mathrm{th}$ International Workshop on Top Quark Physics\\
Durham, UK (videoconference), 14--18 September, 2020
\end{Presented}
\vfill
\end{titlepage}
\def\thefootnote{\fnsymbol{footnote}}
\setcounter{footnote}{0}
%

\section{Introduction}

The associated production of a $t\bar{t}$ pair and a $W^\pm$ gauge boson ($t\bar{t}W$) is one of the heaviest processes that can be studied at the LHC using the full Run II data. For this reason it is an optimal candidate for probing the electroweak (EW) scale by means of precision measurements of top-quark couplings to EW bosons \cite{Dror:2015nkp,Bylund:2016phk}. 
Particularly interesting are the experimental signatures induced by leptonic decays of top quarks and of the $W$ boson, which  can lead to final states with three light leptons or two same-charge light leptons. The latter are relatively rare in the Standard Model (SM) and thus well suited to look for signatures of new physics as predicted by several BSM scenarios, such as Supersymmetry, minimal Universal Extra Dimensions or non-standard Higgs models (see \textit{e.g.} \cite{Barnett:1993ea,Cheng:2002ab,Contino:2008hi}). Other than being interesting in its own right, $t\bar{t}W$ is also an important background for Higgs boson searches in the $t\bar{t}H$ channel as well as for four-top production. Last but not least, the absence of contributions from $gg$-initiated processes up to NNLO in QCD confers distinctively large charge asymmetries to $t\bar{t}W$ in comparison to inclusive $t\bar{t}$ production. Precision studies of these observables offer another opportunity to look for new physics \cite{Maltoni:2014zpa,Frederix:2020jzp,Bevilacqua:2020srb}.

On the experimental side, the ATLAS and CMS Collaborations have carried out direct measurements of the cross section for $t\bar{t}W^\pm$ production at 13 TeV using an integrated luminosity of $\mathcal{L} = 36 \mbox{ fb}^{-1}$ \cite{Sirunyan:2017uzs,Aaboud:2019njj}. More recently, $t\bar{t}W$ has been scrutinized in the context of $t\bar{t}H$ searches in multilepton final states with $\mathcal{L}=80 \mbox{ fb}^{-1}$ \cite{ATLAS:2019nvo} and $\mathcal{L}=137 \mbox{ fb}^{-1}$ \cite{Sirunyan:2020icl}. It also played a role in the first evidence of $t\bar{t}t\bar{t}$ production based on $\mathcal{L}=139 \mbox{ fb}^{-1}$  \cite{Aad:2020klt}. All these analyses report sustained tensions where the measured $t\bar{t}W$ rates exceed SM predictions by $25\%-70\%$ depending on the signal cathegory. This raises the need for increased theoretical accuracy.

On the theory side, several directions have been undertaken in efforts for improving the description of this process at both   inclusive and exclusive levels. This proceedings briefly summarizes the present status and future perspectives of the theoretical modeling of $t\bar{t}W$.

\section{Inclusive $t\bar{t}W^\pm$ production}

When dealing with fully inclusive predictions, the approximation of stable top quarks and $W$ bosons is usually preferred to keep the computational burden of higher-order corrections as small as possible. This strategy allows to achieve predictions for  total production rates with particularly high accuracy. At the present state-of-the-art, the $t\bar{t}W$ cross section is known perturbatively up to the \textit{complete-NLO}, which consists of all possible QCD and EW corrections to the full LO cross section \cite{Hirschi:2011pa,Maltoni:2015ena,Frixione:2015zaa,Frederix:2017wme}. The various NLO contributions are classified into four groups according to the perturbative order: \ordQCDEW{3}{} (dubbed \NLO{1}), \ordQCDEW{2}{2} (\NLO{2}), \ordQCDEW{}{3} (\NLO{3}) and \ordEW{4} (\NLO{4}). By power counting one would naively expect that the contributions above appear in order of decreasing importance, however as shown in Ref.\cite{Frederix:2017wme} a different hierarchy is realized. While \NLO{1} dominates as expected ($\sim 50 \%$ of the LO cross section), \NLO{3} ($\sim 10 \%$) is surprisingly the second most important contribution, followed by \NLO{2} ($\sim -4\%$) and \NLO{4} (permille level). The importance of \NLO{3} is understood as due to the opening of diagrams embedding the  electroweak subprocess of $tW \to tW$ scattering, which is particularly sensitive to the large top-quark Yukawa coupling \cite{Dror:2015nkp,Frederix:2017wme}. 

\begin{figure}[htb]
\includegraphics[height=2.9in]{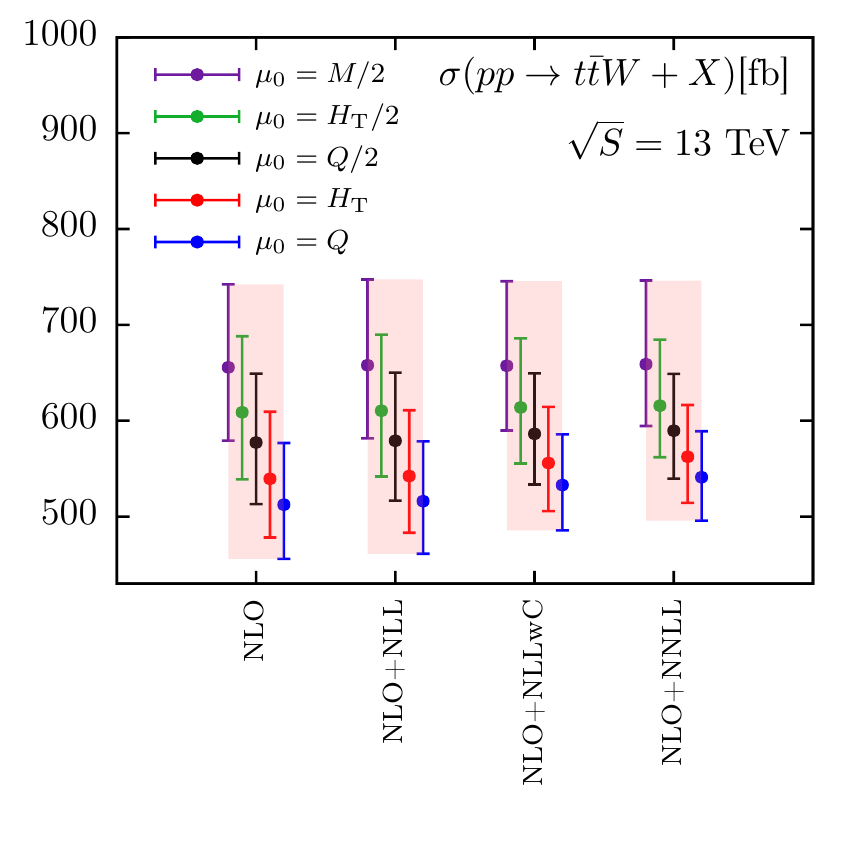}
\put(30,6){
\includegraphics[height=2.7in]{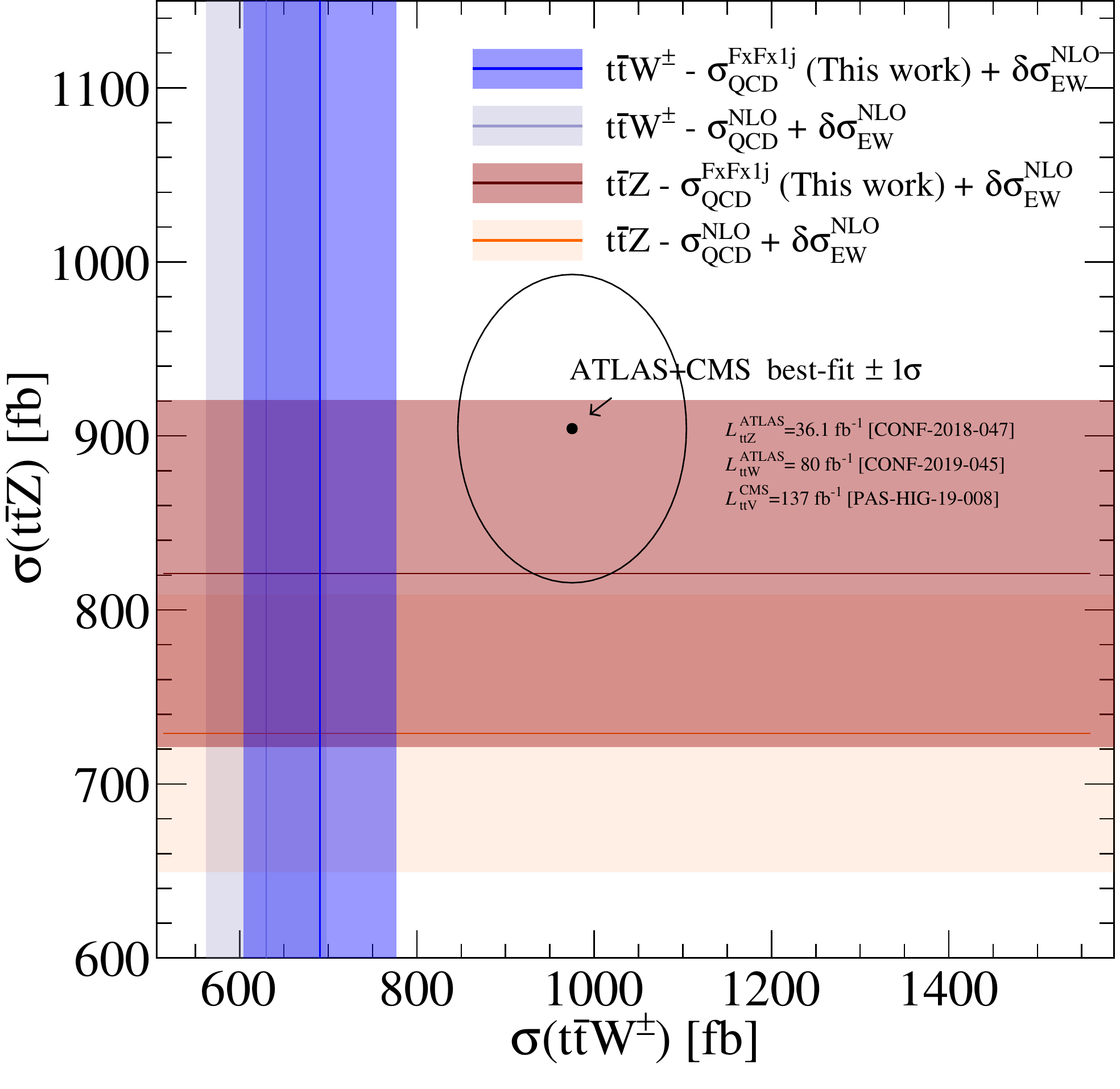}
}
\caption{
\textit{Left plot: impact of soft-gluon resummation on cross section predictions for $t\bar{t}W^\pm$ at $\sqrt{s}=13$ TeV \cite{Kulesza:2020nfh}}. 
\textit{Right plot:} comparison of $t\bar{t}W$ and $t\bar{t}Z$ total cross sections at NLO QCD+EW as well as with FxFx matching \cite{vonBuddenbrock:2020ter}.
}
\label{fig:inclusive_predictions}
\end{figure}

Complete-NLO results have been also matched to threshold resummation of soft gluon corrections up to NNLL accuracy  \cite{Li:2014ula,Broggio:2016zgg,Broggio:2019ewu,Kulesza:2018tqz,Kulesza:2020nfh}, resulting in the most accurate predictions to date for total production rates. After resummation, the $t\bar{t}W$ cross section receives a modest shift of few percents in the central value and no substantial reduction of scale uncertainties, which remain at the level of 10-20$\%$ (see Figure \ref{fig:inclusive_predictions}, left plot). For comparison, in the case of $t\bar{t}Z$ production the central predictions are brought much closer to data and theoretical errors are reduced by almost half \cite{Kulesza:2018tqz}. The weaker impact of resummation on $t\bar{t}W$ is in line with the fact that this process, in contrast with the $gg$-dominated $t\bar{t}Z$, receives only contributions from $q\bar{q}^\prime$ channel at LO. Correspondingly only soft-gluon emissions from incoming quark lines are considered in the resummation frameworks.

To summarize this part, sub-leading EW contributions impact on the $t\bar{t}W$ cross section at the level of $10\%$ and must be necessarily included in experimental analyses. The present state-of-the-art description (NLO+NNLL) does not solve the slight tension observed between theoretical predictions and measured production rates. Much of the attention is on the potential role of missing higher-order corrections, mainly from NNLO QCD. Systematic analyses of $t\bar{t}Wj$ and $t\bar{t}Wjj$ subprocesses, based on NLO multi-jet (FxFx) matching, show that higher jet multiplicities impact the total cross section at the level of $10\%$ \cite{vonBuddenbrock:2020ter,ATLAS:2020esn} (see Figure \ref{fig:inclusive_predictions}, right plot). This raises a strong motivation for a full NNLO QCD calculation of $t\bar{t}W$ which, although highly desirable, remains  beyond reach.

\medskip \medskip

\section{$t\bar{t}W^\pm$ with realistic final states}

Since the lifetime of top quarks and $W$ bosons is extremely short, including effects of decays is an equally important side of the modelling process. Spin correlations should be taken into account as they can have a sizeable impact on the decay products, indeed top-quark polarization effects are particularly important in $t\bar{t}W$ production \cite{Maltoni:2014zpa}.
Also, in view of more accurate comparisons with LHC data, higher-order effects should be included in decays when possible. First attempts in this direction have been made in Ref. \cite{Campbell:2012dh} with the first NLO calculation in full Narrow Width Approximation (NWA), including QCD corrections to both production and decay stages.

At the same time, continuous efforts have been made to improve the modelling of hadronic observables by interfacing  calculations with parton showers. The first study at NLO was presented in Ref. \cite{Garzelli:2012bn} using the POWHEG method as implemented in the \textsc{PowHel} framework. The \NLO{1} accuracy was employed for the production process and LO for spin-correlated decays. Studies based on MC@NLO matching have been also carried out using same perturbative accuracy \cite{Maltoni:2014zpa,Maltoni:2015ena}. In a number of recent papers  \cite{Frederix:2020jzp,Cordero:2021iau,ATLAS:2020esn} the impact of the sub-leading \NLO{3} contributions has been examined in multilepton final states. It is worth to notice that \NLO{3}, like \NLO{1}, can be matched to available shower frameworks such as \textsc{PYTHIA8} since it represents pure QCD corrections to $t\bar{t}W$. In contrast \NLO{2} and \NLO{4} contributions \-- not yet considered at this stage \-- are genuine EW corrections and a consistent matching will require to include EW effects in the shower framework. \NLO{3} effects have not always a flat impact on the phase space and may lead to visible shape distortions. This has been reported \textit{e.g.} for jet multiplicities or dijet invariant mass distributions \cite{Frederix:2020jzp,Cordero:2021iau}. In these cases, including \NLO{3} as a flat $+10\%$ rescaling is clearly not as accurate as including the same effects in a differential manner.

Complementary to parton shower aspects, another direction which can be tackled for improving final-state modelling is to overcome the limit $\Gamma_t/m_t \to 0$ which is inherent in the NWA and characterizes \textit{on-shell} predictions. In \textit{off-shell} calculations, the complete set of resonant and non-resonant diagrams is computed for a given final state (\textit{e.g.} $b\bar{b}e^+\nu_e e^-\bar{\nu}_e\mu^-\nu_\mu +X$ for multilepton channels). Based as it is on a complete calculation at fixed perturbative order, it is also more challenging compared to NWA. Off-shell effects are expected to impact inclusive cross sections at the level of $\mathcal{O}(\Gamma_t/m_t) \approx 0.8\%$ but can reach tens of percents in tails of distributions.
Two independent off-shell calculations of $t\bar{t}W$ in the multilepton channel are currently available at NLO QCD accuracy \cite{Bevilacqua:2020pzy,Bevilacqua:2020srb,Denner:2020hgg}. Very recently further progress has been obtained with the full combination of QCD+EW effects in the same channel \cite{Denner:2021hqi}.

In Ref.\cite{Bevilacqua:2020pzy}, a systematic comparison between off-shell predictions and NWA has been carried out for  different accuracies in the modelling of decays: LO ("$\mbox{NWA}_{\mbox{\scriptsize LOdecay}}$") and NLO QCD ("full NWA"). Inclusive fiducial cross sections show a very good agreement: off-shell results differ from $\mbox{NWA}_{\mbox{\scriptsize LOdecay}}$ by 3-5\% and just by 0.2\% from the full NWA. These numbers should be compared to theoretical uncertainties estimated from scale variation, which amount to $11\%$ ($\mbox{NWA}_{\mbox{\scriptsize LOdecay}}$), $7\%$ (full NWA) and $7\%$ (off-shell). Interestingly, the accuracy of decays is found to impact scale uncertainties in NWA.  
Much larger discrepancies are observed between off-shell predictions and NWA when looking at differential cross sections: as an example we highlight the case of the transverse momentum ($p_T$) of the leading $b$-jet. The tail of the distribution is enhanced up to 50\% by off-shell effects, to be compared with $\sim 20\%$ scale uncertainties in the same region \cite{Bevilacqua:2020pzy}.  An alternative approach for assessing non-resonant effects is discussed in Ref.\cite{Denner:2020hgg}, where Double Pole Approximation (DPA) is used for the virtual and for the $I$-operator contributions to the NLO cross section. Comparing with full off-shell results, the fiducial cross-section agrees at percent-level while discrepancies up to 10\% are found in regions not dominated by the $t\bar{t}$ resonance.
Both Refs.\cite{Bevilacqua:2020pzy,Denner:2020hgg} propose dynamical scales which help to obtain moderate QCD corrections and reduced shape distortions  for most observables. There are notable exceptions, \textit{e.g.} the $p_T$ of the $b\bar{b}$ system, where QCD corrections reach up to 300\% in suppressed phase space regions \cite{Denner:2020hgg}. The large $K$-factors are consistent with the dominance of effects from real radiation, which is also confirmed by the observation of LO-like scale uncertainties in the same regions.

\section{Conclusions}

Sustained tensions with measurements of enhanced production rates at the LHC raise the need for improved theoretical modeling of the $t\bar{t}W$ process. The state-of-the-art accuracy for inclusive predictions is NLO+NNLL. On a more exclusive ground, parton-shower Monte Carlo generators can now model on-shell $t\bar{t}W$ production including EW effects at \ordQCDEW{}{3}. Additionally, the impact of off-shell and non-resonant effects in $t\bar{t}W$ multilepton final states can be studied by means of complete fixed-order calculations. 

Achieving full NNLO QCD predictions, while beyond  present reach, will be an important step forward to reduce theoretical uncertainties as well as to assess the possible role of higher-order effects in the observed tension with data. In view of accurate comparisons with high-luminosity data, it will be also appropriate to include the full set of sub-leading EW effects in Monte Carlo generators. Matching off-shell calculations to parton showers is an ambitious goal which will lead to the most complete NLO description of $t\bar{t}W$ final states.


\end{document}